\documentclass[10pt, journal, paper, twocolumn]{IEEEtran}
\usepackage[noadjust]{cite}      
\usepackage{graphicx}  
\usepackage{amsmath,amssymb,epsfig}
\usepackage{multicol}
\usepackage{algorithm}
\usepackage{url}
\usepackage{float}
\usepackage{amsfonts}
\usepackage{theorem}

\makeatletter
    \newcommand\figcaption{\def\@captype{figure}\caption}
    \newcommand\tabcaption{\def\@captype{table}\caption}
\makeatother

\newtheorem{defn}{Definition}

\begin{document}
\title{Iterative Soft-Input-Soft-Output Decoding of Reed-Solomon Codes by
Adapting the Parity Check Matrix \thanks{Jing~Jiang and Krishna~R.~Narayanan are with Department of Electrical
and Computer Engineering Texas A\&M University, College Station, TX 77843, USA, (email: jjiang@ece.tamu.edu;
krn@ece.tamu.edu). This work was supported in part by the National Science Foundation under grant CCR0093020 and
by Seagate Technology, Pittsburgh, PA 15222 USA}}
\author{Jing~Jiang~\IEEEmembership{Student Member,~IEEE,} and Krishna~R.~Narayanan~\IEEEmembership{Member,~IEEE,}\\}
\maketitle{}

\begin{abstract}
An iterative algorithm is presented for soft-input-soft-output
(SISO) decoding of Reed-Solomon (RS) codes. The proposed iterative
algorithm uses the sum product algorithm (SPA) in conjunction with
a binary parity check matrix of the RS code. The novelty is in
reducing a submatrix of the binary parity check matrix that
corresponds to less reliable bits to a sparse nature before the
SPA is applied at each iteration. The proposed algorithm can be
geometrically interpreted as a two-stage gradient descent with an
adaptive potential function. This adaptive procedure is crucial to
the convergence behavior of the gradient descent algorithm and,
therefore, significantly improves the performance. Simulation
results show that the proposed decoding algorithm and its
variations provide significant gain over hard decision decoding
(HDD) and compare favorably with other popular soft decision
decoding methods.
\end{abstract}


\section{Introduction}
\label{sec:introduction}

Reed-Solomon (RS) codes are one of the most popular error
correction codes in many state-of-the-art communication and
recording systems. In most of these existing systems, RS codes are
decoded via an algebraic hard decision decoding (HDD) algorithm.
When soft information about the channel output is available, HDD
can incur a significant performance loss compared to optimal soft
decision decoding.  For example, for the AWGN channel, the loss is
believed to be about 2-3~dB. Moreover, in some situations, it is
desirable to obtain soft output from the decoder. A typical
example is when turbo equalization is employed at the receiver and
soft outputs from the decoder have to be fedback to the equalizer.
Consequently, soft-input-soft-output (SISO) decoding algorithms
for RS codes are of research interest both for theoretical and
practical reasons.

In the literature, there are several classes of soft decision decoding algorithms. Enhanced HDD algorithms such
as generalized minimum distance (GMD) decoding \cite{forney_gmd}, Chase decoding \cite{chase_chase} and a hybrid
of Chase and GMD algorithms (CGA) \cite{tang_cga} use reliability information to assist HDD decoding. Enhanced
HDD usually gives a moderate performance improvement over HDD with reasonable complexity. Recently, algebraic
soft interpolation based decoding (the Koetter-Vardy (KV) algorithm \cite{koetter_kv}), which is a list decoding
technique that uses the soft information from the channel to interpolate each symbol, has become popular
\cite{el-khamy_kv} \cite{nayak_kv} \cite{gross_vlsi}.  The KV algorithm can significantly outperform HDD for low
rate RS codes. However, to achieve large coding gain, the complexity can be prohibitively large. For detailed
discussions of the complexity performance tradeoff of the KV algorithm, we refer interested readers to
\cite{gross_vlsi}.  Another approach is decoding RS codes using their binary image expansions. Vardy and Be'ery
showed that RS codes can be decomposed into BCH subfield subcodes which are glued together using glue vectors
\cite{vardy_rs}. Even though this decomposition significantly reduces the trellis complexity of maximum
likelihood (ML) decoding of RS codes, the complexity still grows exponentially with the code length and
$d_{min}$ and it is thus infeasible for practical long codes. Recent work \cite{ponn_rs} reduces the complexity
and modifies the algorithm in \cite{vardy_rs} to generate soft output efficiently. By using the binary image
expansion of RS codes, we can also use decoding algorithms for general linear block codes such as reliability
based ordered statistics decoding (OSD) \cite{fossorier_osd} and its variations \cite{yingquan_osd} for soft
decision decoding of RS codes. Previous such works include the hybrid algorithm by Hu and Shu Lin \cite{hu_osd}
and the box and match algorithm (BMA) \cite{fossorier_bma} by Fossorier and Valembois. OSD based algorithms are
quite efficient for practical RS codes even though they do not take the structure of the RS codes into account.

Iterative decoding \cite{hagenauer_app} algorithms are of emerging
interest for soft decision decoding of RS codes
\cite{ungerboeck_rs}, \cite{jiang_ssid}, \cite{yedidia_gfft}. The
main difficulty in directly applying iterative decoding techniques
to RS codes is that the parity check matrix of an RS code is in
general not sparse.  In order to deal with such dense parity check
matrices, Yedidia $et~al.$ \cite{yedidia_gbp} proposed a
``generalized belief propagation'' (GBP) algorithm that introduces
hidden states in iterative decoding. However, their results show
that this technique does not work well for high density parity
check (HDPC) codes (such as RS codes) over the AWGN channel. We
observe from the simulations that the iterative decoder fails
mainly due to some of the unreliable bits ``saturating'' most of
the checks which causes iterative decoding to be stuck at some
pseudo-equilibrium points. In \cite{jiang_ssid}, the cyclic
structure of RS codes is taken advantage of and a sum product
algorithm (SPA) is applied to a random shift of the received
vector at each iteration to avoid pseudo-equilibrium points (see
\cite{jiang_ssid} for details). While significant improvement in
performance over HDD was obtained for short codes, the performance
improvement diminishes for long RS codes.

In this paper, we present an iterative SISO decoding algorithm (which is based on the SPA) for RS codes. The
main novelty in the proposed scheme is to adapt the parity check matrix at each iteration according to the bit
reliabilities such that the unreliable bits correspond to a sparse submatrix and the SPA is then applied to the
adapted parity check matrix. This adaptation prevents the iterative decoder from getting stuck at
pseudo-equilibrium points and, hence, the convergence behavior of the iterative decoder is significantly
improved. Simulation results show that the proposed iterative decoding scheme performs well for RS codes with
reasonable decoding complexity, even though the parity check matrices are not sparse. While the approach in
\cite{jiang_ssid} is also one of adapting the parity check matrix, the adaptation there is restricted to be
within the class of cyclic shifts of the parity check matrix, whereas we consider a more general adaptation
procedure here which is based on the bit reliabilities. The proposed algorithm can be applied to any linear
block code; however, we restrict our attention to RS codes in this paper because of the practical interest in
soft decision decoding of RS codes and the fact that the gain from this adaptive procedure is significant for
codes with dense parity check matrices such as RS codes.

The rest of the paper is organized as follows: The generic
iterative decoding algorithm is presented in Section
\ref{sec:algorithm}. A geometric interpretation of the proposed
algorithm is given in Section \ref{sec:geometry}.  Several
variations of the generic algorithm are investigated in Section
\ref{sec:variations}. In Section \ref{sec:sim}, simulation results
of the proposed algorithm are presented and compared with popular
RS soft decoding algorithms. Discussions and conclusions are
presented in Section \ref{sec:conclusion}.

\section{Iterative Decoding Algorithm by Adapting the Parity Check Matrix}
\label{sec:algorithm}

Consider a narrow sense $(N,K)$ RS code over $GF(2^m)$ which has a
minimum distance $d_{min} = \delta = N-K+1$. The parity check
matrix over $GF(2^m)$ can be represented as follows:

\begin{equation} \label{eq:prt_matrix}
   \textbf{H}_{s}=\left(\begin {array}{cccccc}
                     1 &  \beta   & \cdots & \beta^{(N-1)}\\
                     1 &  \beta^2 & \cdots & \beta^{2(N-1)}\\
                       &          & \cdots &        \\
                     1 &  \beta^{(\delta-1)} & \cdots &
\beta^{(\delta-1)(N-1)}\\
              \end{array} \right)\;
 \end{equation}
where $\beta$ is a primitive element in $GF(2^m)$.  Let $n = N \times m$ and  $k = K \times m$ be the length of
the codeword and the information at the bit level, respectively. $\textbf{H}_{s}$ has an equivalent binary image
expansion $\textbf{H}_{b}$ (see \cite{lin_book} for details), where $\textbf{H}_{b}$ is an $(n-k) \times n$
binary parity check matrix.

We will use underlined letters to denote vectors and bold face letters to denote matrices. Let
$\underline{c}~=~[c_1,c_2,\ldots,c_{n}]$ be the binary representation of an RS codeword. In the description of
the generic algorithm, we first assume that the bits are modulated using BPSK (with $0$ mapped to $+1$ and $1$
mapped to $-1$) and transmitted over an AWGN channel (extension to other channels is straightforward). The
received vector is given by

\begin{equation}
\centering \label{eqn:awgn} \underline{y} = (-2\underline{c}+1)+\underline{n},
\end{equation}
Thus, the initial reliability of each bit in the received vector can be expressed in terms of the log-likelihood
ratios (LLR) as observed from the channel:

\begin{equation}
\centering \label{eqn:llr2} \emph{L}^{(0)}(c_i) = \log{\frac{P(c_i=0|y_i)}{P(c_i=1|y_i)}},
\end{equation}

\begin{figure}
\begin{center}
\includegraphics[width=3.0in]{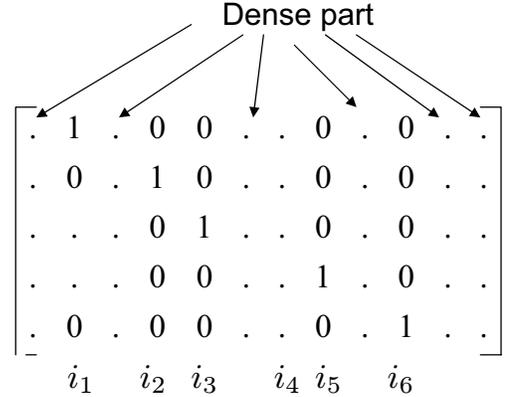}
\caption{Form of the Parity Check Matrix Suitable for Iterative Decoding Obtained through Row Operations}
\label{matrixform}
\end{center}
\end{figure}

The proposed algorithm is composed of two stages: the matrix updating stage and the bit-reliability updating
stage. In the matrix updating stage, the magnitude of the received LLR's $|\emph{L}(c_i)|$ are first sorted and
let $i_1,i_2, \ldots,i_{N-K}, \ldots,i_{n}$ denote the position of the bits in terms of ascending order of
$|\emph{L}(c_i)|$, i.e., the bit $c_{i_1}$ is the least reliable and $c_{i_{n}}$ is the most reliable. We begin
with the original parity check matrix \textbf{H$_b$} and first reduce the $i_1^{th}$ column of \textbf{H$_b$} to
a form $[1 \ 0 \ldots 0]^T$. Then, we reduce the $i_2^{th}$ column of \textbf{H$_b$} to a form $[0 \ 1 \ 0
\ldots 0]^T$ and so on. We can be guaranteed to proceed until the $i_{(N-K)}^{th}$ column, since there are at
least $(N-K)$ independent columns in \textbf{H$_b$}. Then we try to reduce the $i_{N-K+1}^{th}$ column to $[
\underbrace{0 \ldots 0 }_{(N-K)} 1, 0, \ldots, 0]^T$. However, there is no guarantee we can do this. If we are
unable to do so, we will leave that particular column and try to reduce $i_{(N-K+2)}^{th}$ column to the above
form and so on. Finally, we can reduce $(n-k)$ columns among the $n$ columns of \textbf{H$_b$} to be the
identity matrix, since the matrix is $(n-k) \times n$ and is full rank. The matrix is thus reduced to a form as
shown in Fig.~\ref{matrixform}. We denote the set of unreliable bits corresponding to the sparse submatrix as
$\emph{\underline{B}}_L$.

The proposed algorithm is iterative and during the $l^{th}$
iteration, we have a vector of LLR's as:

\begin{equation}\label{eqn:llr vector}
\emph{\underline{L}}^{(l)} = [L^{(l)}(c_1), L^{(l)}(c_2), \cdots, L^{(l)}(c_n)]
\end{equation}
where initially $\emph{\underline{L}}^{(0)}$ is determined from the channel output. Then, the parity check
matrix is reduced to a desired form based on $\emph{\underline{L}}^{(l)}$:

\begin{equation}\label{matrix_update}
\textbf{H}_b^{(l)} = \phi(\textbf{H}_b,|\emph{\underline{L}}^{(l)}|).
\end{equation}
In the bit-reliability updating stage, the extrinsic LLR vector $\emph{\underline{L}}^{(l)}_{ext}$ is first
generated according to $\emph{\underline{L}}^{(l)}$ using the SPA \cite{hagenauer_app} based on the adapted
parity check matrix $\textbf{H}_b^{(l)}$:

\begin{equation}\label{llr_ext}
    \emph{\underline{L}}^{(l)}_{ext} =
\psi(\textbf{H}_b^{(l)},\emph{\underline{L}}^{(l)})
\end{equation}

That is for each bit, the extrinsic LLR is updated according to:
\begin{equation}\label{ext_update}
\emph{L}^{(l)}_{ext}(c_i) = \mathop{\sum_{j=1}^{n-k}}_{H^{(l)}_{ji} = 1}{2\tanh^{-1} \left( \mathop{\prod_{p =
1}^{n}}_{p \neq i, H^{(l)}_{jp} = 1}{\tanh \left(\frac{\emph{L}^{(l)}(c_p)}{2}\right)} \right) }
\end{equation}

The bit-reliability is then updated as:
\begin{equation}\label{llr_update}
    \emph{\underline{L}}^{(l+1)} = \emph{\underline{L}}^{(l)}+\alpha
\emph{\underline{L}}_{ext}^{(l)}
\end{equation}
where $0 < \alpha \leq 1$ is a damping coefficient. This is continued until a predetermined number of times
$l_{max}=N_1$ or until all the parity checks are satisfied. A detailed description of the algorithm is given in
Algorithm \ref{alg:generic}.

\begin{algorithm}
\begin{description}
\item[\textbf{Step1.}] Initialization: set $\alpha$, $l_{max}=N_1$, $l = 0$ and the LLR's for the coded bits
from the channel observation: $\underline{L}^{(0)} = \frac{2}{\sigma^2}\underline{y}$

\item[\textbf{Step2.}] Reliability  based parity check matrix adaptation: $\textbf{H}_b^{(l)} =
\phi(\textbf{H}_b,|\emph{\underline{L}}^{(l)}|)$.

    a) Order the coded bits according to the absolute value of the
    LLR's $|\emph{\underline{L}}^{(l)}|$ and record the
    ordering indices.

    b) Implement Gaussian elimination to systematize the $(n-k)$
    unreliable positions which are independent in the parity check
    matrix. (The submatrix can also be made to be degree-2
    connected, see Section \ref{subsec:deg2}).

    \item[\textbf{Step3.}] Extrinsic information generation:
        Apply SPA to generate the extrinsic LLR for
each bit using the adapted parity check matrix $\textbf{H}_b^{(l)}$:

$\emph{\underline{L}}_{ext}^{(l)} = \psi(\textbf{H}_b^{(l)},\emph{\underline{L}}^{(l)})$ (according to
(\ref{ext_update})).

    \item[\textbf{Step4.}] Bit-level reliabilities update:

            $\emph{\underline{L}}^{(l+1)} = \emph{\underline{L}}^{(l)}+\alpha
            \emph{\underline{L}}_{ext}^{(l)}$, where $0 < \alpha \le 1$.

   \item[\textbf{Step5.}] Hard decision:
        $\hat{c_i}=\left\{%
        \begin{array}{ll}
            0, & \hbox{$L^{(l+1)}(c_i)>0$;}\\
            1, & \hbox{$L^{(l+1)}(c_i)<0$.}\\
        \end{array}%
        \right.$

    \item[\textbf{Step6.}] Termination criterion:
        If all the checks are satisfied, output the estimated bits;
        else if $l = l_{max}$, declare a decoding failure;
        otherwise set $l \leftarrow l+1$ and go to \textbf{Step2} for
another iteration.
\end{description}
\caption{Iterative Decoding Algorithm by Adapting the Parity Check Matrix} \label{alg:generic}
\end{algorithm}

The proposed adaptive algorithm is inspired by the OSD \cite{fossorier_osd}. However, instead of reprocessing
the most reliable basis (MRB), we adapt the parity check matrix according to the least reliable basis (LRB). It
can also be viewed as a generalization of the iterative \emph{a posteriori} probability (APP) decoding algorithm
based on a set of minimum weight parity check vectors by Lucas \emph{et al.} \cite{lucas_appdec}. In
\cite{lucas_appdec}, the iterative algorithm is interpreted as a gradient descent. The adaptive algorithm
generalizes the idea of gradient descent and extends it to be a two-stage gradient descent algorithm with an
adaptive potential function. The damping coefficient $\alpha$ serves as the step size in the gradient descent
process to control the dynamics of convergence. In the following section, we look into a geometric
interpretation of this algorithm.

\section{Geometric Interpretation of the Proposed Algorithm}
\label{sec:geometry}

In this section, a geometric interpretation of the proposed algorithm as a two-stage optimization procedure is
presented. The idea of using optimization methods, such as gradient descent, to solve decoding problems can be
dated back to Farrell \emph{et al.} \cite{farrell_opt}. The belief propagation (BP) based algorithms by Gallager
\cite{gallager63} and Pearl \cite{pearl98} were also shown to be special cases of the gradient descent
algorithm. The bit reliability updating algorithm in this paper is more similar to that proposed by Lucas
\emph{et al.} in \cite{lucas_appdec}.

Define the operator $\nu: [-\infty,+\infty]\rightarrow[-1,1]$ as a
mapping from the LLR domain to $\tanh$ domain:
\begin{equation}\label{eqn:tanh_map}
\nu(L) = \tanh \left(\frac{L}{2}\right) = \frac{e^L-1}{e^L+1}
\end{equation}
where the mapping is one-to-one and onto.

It is immediate that the inverse operator $\nu^{-1}:
[-1,+1]\rightarrow[-\infty,+\infty]$ can be expressed as:
\begin{equation}\label{eqn:ln_map}
\nu^{-1}(t) = \ln \left( \frac{1+t}{1-t} \right),~~~~~t\in[-1,+1]
\end{equation}

We apply the one-to-one $\tanh$ transform on the LLR's and get the
reliability measure of the received signal in the $\tanh$ domain
as:
\begin{equation}\label{eqn:tanh vector}
\emph{\underline{T}} = [T_1, T_2, \cdots, T_{n}] = [\nu(L(c_1)), \cdots, \nu(L(c_n))]
\end{equation}
As in \cite{hagenauer_app}, we can measure the reliability of the
$j^{th}$ parity check node as:
\begin{equation}\label{eqn:tanh_llr}
\gamma_j = \mathop{\prod_{p = 1}^{n}}_{H_{jp} = 1}{\nu(L(c_p))}
\end{equation}

Following the concept of a generalized weighted syndrome proposed by Lucas \emph{et al.} (Eqn. (20) in
\cite{lucas_appdec}), we define a cost function $J$, which characterizes the reliability of the received vector
$\emph{\underline{T}}$ with a particular parity check matrix $\textbf{H}_b$.
\begin{defn} \label{defn:soft syndrome sum}
Define the potential function $J$ as:
\begin{equation}\label{eqn:soft syndrome sum}
J(\textbf{H}_{b},\emph{\underline{T}}) = -\sum_{j=1}^{(n-k)}{\gamma_j} = -\sum_{j=1}^{(n-k)}{\mathop{\prod_{p =
1}^{n}}_{H_{jp}=1}{T_p}}
\end{equation}
\end{defn}
where $J$ is a function of both the parity check matrix $\textbf{H}_{b}$ and the received soft information
$\emph{\underline{T}}$.

The operator $\nu$ maps the original $n$-dimensional unbounded
real space into an $n$-dimensional cube (since the output of the
$\tanh$ function is confined to [-1, 1]). The potential function
$J$ is minimized iff a valid codeword is reached, that is all the
checks are satisfied and $|T_j| = 1$ for $j = 1, \cdots, n$, where
$J_{min} = -(n-k)$. Besides, points with all $|T_j| = 1$
correspond to vertices of the $n$-dimensional cube. Therefore,
valid codewords correspond to the vertices of the $n$-dimensional
cube at which the potential function has the minimum value of
$-(n-k)$. The decoding problem can be interpreted as searching for
the most probable minimum potential vertex given the initial point
observed from the channel.

Note that the potential function $J$ is minimized iff a valid
codeword is reached. It is quite natural to apply the gradient
descent algorithm to search for the minimum potential vertex, with
the initial value \emph{\underline{T}} observed from the channel.
Consider the gradient of $J$ with respect to the received vector
$\emph{\underline{T}}$. From (\ref{eqn:soft syndrome sum}), it can
be seen that:
\begin{equation}\label{eqn:soft syndrome gradient}
\nabla J(\textbf{H}_{b},\emph{\underline{T}}) =
\left(\frac{\partial{J}(\textbf{H}_{b},\emph{\underline{T}})}{\partial{T_1}},
\frac{\partial{J}(\textbf{H}_{b},\emph{\underline{T}})}{\partial{T_2}},
\cdots,
\frac{\partial{J}(\textbf{H}_{b},\emph{\underline{T}})}{\partial{T_{n}}}\right)
\end{equation}
where the component wise partial derivative with respect to $T_i$ is given by:
\begin{equation}\label{eqn:partial derivative}
\frac{\partial{J}(\textbf{H}_{b},\emph{\underline{T}})}{\partial{T_i}} =
-\mathop{\sum_{j=1}^{(n-k)}}_{H_{ji}=1}{\mathop{\prod_{p = 1}^{n}}_{p \neq i, H_{jp}=1}{T_p}}
\end{equation}
Thus, the gradient descent updating rule can be written as:
\begin{equation}\label{eqn:gradient update}
\emph{\underline{T}}^{(l+1)} \leftarrow \emph{\underline{T}}^{(l)}- \alpha \nabla
J(\textbf{H}_{b}^{(l)},\emph{\underline{T}}^{(l)})
\end{equation}
where $\alpha$ is a damping coefficient as in standard gradient descent algorithms to control the step width.

Note that the reliabilities in $\tanh$ domain are confined to
$T_i\in [-1,1]$. However, the updating rule (\ref{eqn:gradient
update}) does not guarantee this. Therefore, we use the following
modified updating rule to guarantee that the updated $T_i$'s $\in
[-1,1]$:
\begin{equation}\label{eqn:llr gradient update}
T_i^{(l+1)} \leftarrow \nu \left( \nu^{-1}\left(T_i^{(l)}\right)- \alpha \left[
-\sum_{H^{(l)}_{ji}=1}{\nu^{-1}\left(\prod_{p \neq i, H^{(l)}_{jp} = 1}{T_p^{(l)}}\right)}\right]\right)
\end{equation}
where $\nu^{-1}(x) = 2 \tanh^{-1}(x)$. It can be seen that the
above non-linear updating rule is exactly the same as Step 3-Step
4 in Algorithm \ref{alg:generic}.

When iterative decoding is applied to an HDPC code, with very high
probability, the iterative algorithm will reach some local minimum
points where $\nabla J(\textbf{H}_{b},\emph{\underline{T}})$ is
zero or is close to zero (since a few unreliable symbols will
render the components of $\nabla
J(\textbf{H}_{b},\emph{\underline{T}})$ to be small or close to
zero). We refer to these as pseudo-equilibrium points since
gradient descent gets stuck at these points while these points do
not correspond to valid codewords.

From (\ref{eqn:soft syndrome sum}), we observe that since $J$ is
also a function of $\textbf{H}_{b}$, different choices of the
parity check matrices $\textbf{H}_{b}$, though span the same dual
space, result in different potential functions $J$. More
importantly, each $\textbf{H}_{b}$ results in a different gradient
$\nabla J(\textbf{H}_{b},\emph{\underline{T}})$. The proposed
algorithm exploits this fact and when a pseudo equilibrium point
is reached, by adapting the parity check matrix based on the bit
reliabilities, switches to another $\textbf{H}_{b}$ such that it
allows the update in (\ref{eqn:llr gradient update}) to proceed
rather than getting stuck at the pseudo-equilibrium point.
However, note that the potential function that we want to minimize
does not involve the Euclidean distance between the received
codeword and current estimate at all. That is, the adaptive
algorithm attempts merely to find a codeword that satisfies all
the parity checks, without really enforcing that it be the one at
minimum distance from the received word. Since small steps are
taken in the gradient descent, very often we converge to the
codeword at small distance from the received vector as well.
However, there is no guarantee of convergence to the nearest
codeword.

We use the following examples to show the operation of the adaptive algorithm and its difference from directly
applying iterative decoding to an HDPC code.

\underline{Example~1:} Consider decoding of a random linear block
code where each entry of the parity check matrix is i.i.d and 0 or
1 with equal probability over an erasure channel. We first apply
the gradient descent algorithm directly to HDPC codes without
reliability based adaptation. Assume that the erasure fraction is
$\epsilon$, therefore the number of erased bits is $n \epsilon$.
Consider a particular parity check, any code bit will participate
in that check with probability 1/2 (according to the i.i.d.
equiprobable assumption). A check is not erased iff all the
participated bits are not erased. Therefore, the probability that
the $s^{th}$ check is erased is

\begin{equation}\label{eqn:bec_no_adaptation}
Pr\{s^{th} \text {check is erased} \} \doteq 1-(\frac{1}{2})^{n\epsilon}, n \rightarrow \infty
\end{equation}
Assume that the $i^{th}$ bit is erased and it participates in $r$ parity checks in the parity check matrix. The
$i^{th}$ component of the gradient vector is zero (i.e.,
$\frac{\partial{J}(\textbf{H}_{b},\emph{\underline{T}})}{\partial{T_i}}=0$) iff the extrinsic LLR's from all the
checks it participates in are erased. The probability that the $i^{th}$ component of the gradient is zero is:
\begin{align}\label{align:prob_erasure}
\nonumber Pr\{\frac{\partial{J}(\textbf{H}_{b},\emph{\underline{T}})}{\partial{T_i}}
= 0 \}  &\doteq [1-2^{-(n\epsilon-1)}]^r\\
\nonumber&\doteq 1-r2^{-(n\epsilon-1)}+o(2^{-(n\epsilon-1)})\\
&\doteq 1-r2^{-(n\epsilon-1)}\rightarrow 1, n \rightarrow \infty
\end{align}
which suggests that unless the number of parity checks grows
exponentially with $n \epsilon$, iterative decoding gets stuck at
a pseudo-equilibrium point with high probability.

On the other hand, for the BEC, it is known that by adapting the parity check matrix corresponding to less
reliable bits (i.e. the erased bits), ML decoding performance can be achieved \cite{elias_bec} in one iteration.
If the Gaussian elimination is successful, then all the erasures can be recovered. Gaussian elimination will not
be successful iff some of the columns corresponding to the erased bits are dependent. In this case, there is
ambiguity between two or more valid codewords. That is, the ML decoder also fails.

In conclusion, for the BEC, gradient descent without adaptation
tends to get stuck at a pseudo-equilibrium point, while the
reliability based adaptation will help gradient descent to
converge to the ML solution in one iteration.

\begin{figure}
\begin{center}
\includegraphics[width=3.0in]{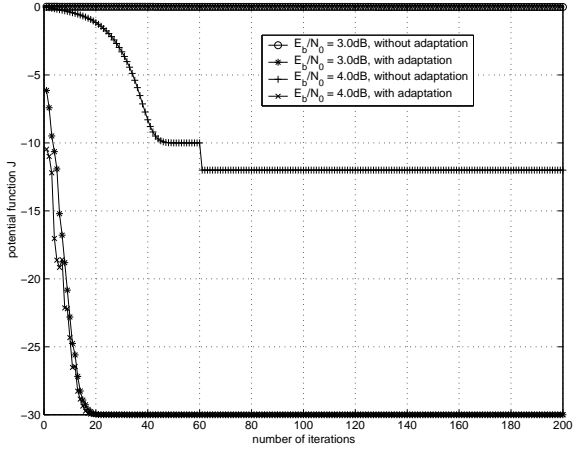}
\caption{Convergence Behavior of Iterative Decoding with and without Adaptation of RS(31, 25)}
\label{fig:convergence}
\end{center}
\end{figure}

\underline{Example~2:} The idea of reliability based parity check
matrix adaptation can naturally be extended to AWGN channels and
the insight remains the same. Though adapting the parity check
matrix based on the channel output does not guarantee convergence
to the ML decision for AWGN channels, it does avoid iterative
decoding getting stuck at pseudo-equilibrium points and thus
improves the convergence behavior. We give a numerical example of
the convergence behavior of iterative decoding of an RS(31,25)
code in Figure \ref{fig:convergence}. A typical realization of
iterative decoding is simulated. The potential function $J$ is
plotted against the number of iterations. Since there are 30
parity checks for RS(31, 25), the minimum value of the potential
function is $J = -30$ (corresponding to valid codewords). We can
see that due to the high density of the parity check matrix of the
RS code, iterative decoding without matrix adaptation (Algorithm
\ref{alg:generic} without Step 2) gets stuck at some
pseudo-equilibrium. On the other hand, when the iterative
algorithm is applied in conjunction with reliability based parity
check matrix adaptation (Algorithm \ref{alg:generic}), the value
of $J$ quickly goes to the global minimum as the number of
iteration increases. Consequently, reliability based parity check
matrix adaptation improves the convergence behavior of iterative
decoding significantly. We will show in Section \ref{sec:sim} that
the adaptive algorithm also significantly improves the error
performance.

\section{Variations to the Generic Algorithms}
\label{sec:variations}

In this section, several variations of the proposed algorithm are
discussed either to improve the performance or to reduce the
decoding complexity.

\subsection{Degree-2 Random Connection}
\label{subsec:deg2}

One problem with the proposed approach is that since each bit in the unreliable part $\emph{\underline{B}}_L$
participates in only one check, it receives extrinsic information from one check only. If there is a bit error
in the dense part participating in that check, the bit in $\emph{\underline{B}}_L$ tends to be flipped and the
decoder tends to converge to a wrong codeword. To overcome this drawback, we can reduce the matrix
$\textbf{H}_b$ to a form where the submatrix corresponding to the less reliable bits is sparse (say column
weight 2 rather than 1). This can improve the performance since each less reliable bit now receives more
extrinsic information while the submatrix corresponding to the unreliable bits still does not form any loops
(i.e., there is no loop involving only the unreliable bits). We can obtain this via a degree-2 random connection
algorithm. The details are presented in Algorithm \ref{alg:deg2}.

\begin{algorithm}
\begin{description}
    \item[\textbf{Step1.}] Apply Gaussian elimination to the parity check matrix
    and obtain an identity matrix in the unreliable part.

    \item[\textbf{Step2.}] Generate a random permutation of numbers from 1 to
    n-k. Record all the indices, i.e., $p_1,p_2,p_3,\cdots,p_{n-k}$.

    \item[\textbf{Step3.}] Adapt the parity check matrix
    according to the follow rule: add $p_{i+1}^{th}$ row to $p_{i}^{th}$ row,
    for i = 1 to n-k-1.
\end{description}
\caption{Deg-2 random connection algorithm} \label{alg:deg2}
\end{algorithm}

After the Deg-2 random connection, all the (n-k-1) columns in the parity check matrix are of degree-2 except the
$p_1^{th}$ column. The last column $p_1$ can be left of degree-1, which will not significantly affect the
performance. This appears to improve the performance of the proposed algorithm especially in the high SNR's.

\subsection{Various Groupings of Unreliable Bits}
\label{subsec:grouping}

Another variation that can help to further improve the performance is to run the proposed algorithm several
times each time with the same initial LLR's from the channel but a different grouping of the less reliable bits.
It is possible that some bits with $|L^{(l)}(c_i)|$ close to those in the unreliable set
$\emph{\underline{B}}_L$ are also of the wrong sign and vice-versa. Hence, we can run the proposed algorithm
several times each time with a different grouping of the less reliable bits. That is, we can swap some bits in
the reliable part with those in the unreliable part near the boundary and run the matrix adaptation all over
again, which gives a new $\textbf{H}_b$. We then run the proposed algorithm on that new matrix $\textbf{H}_b$.
Each time the proposed algorithm is run, a different estimate of codeword may be obtained due to the difference
in the parity check matrix $\textbf{H}_b$. All the returned codewords are kept in a list and finally the one
that minimizes Euclidean distance from the received vector is chosen. We will see from simulation results that
this method can significantly improve the asymptotic performance, but also increases the worst case complexity.
Similar techniques have been used in conjunction with OSD by Fossorier \cite{fossorier_iist} and Wu
\cite{yingquan_osd}. The way of grouping reliable bits used here is similar to the grouping scheme by Wu
\cite{yingquan_osd}. We refer interested readers to \cite{yingquan_osd} for a detailed description and
asymptotic performance analysis.

\subsection{Incorporated Hard Decision Decoding}
\label{subsec:hdd}

A hard decision decoder can be used during each iteration in the proposed algorithm to improve the performance
and accelerate decoding as well.  Since the HDD may return a codeword which is different from the ML codeword,
we do not stop the decoder once a codeword is returned by the HDD. Rather, we still iterate up to a maximum
number of iterations to obtain all the codewords returned by HDD during each iteration and finally pick up the
most likely codeword. This guarantees to perform no worse than the proposed algorithm or HDD. In practice, error
detection schemes such as a cyclic redundancy check (CRC) or other test techniques as discussed in
\cite{fossorier_jsac} can serve as a practical stopping criterion to reduce the average decoding complexity.
Combining the adaptive scheme with other SIHO algorithms such as the KV algorithm has recently been investigated
in \cite{el-khamy_hybrid}.

\subsection{Partial Reliable Bits Updating}
\label{subsec:partial}

The complexity in the bit level reliabilities update part can be further reduced via ``partial reliable bits
updating'' scheme. The main floating-point operation complexity comes from the computation of the extrinsic
information in the reliable part (where the submatrix is dense). However, in the adaptation of the parity check
matrix, only some bits in the boundary will be switched from the reliable part to the unreliable part.
Therefore, in the bit reliability updating stage, we only update the bits in the unreliable set
$\emph{\underline{B}}_L$ and some reliable bits with $|L^{(l)}(c_j)|$ close to those in the unreliable set
$\emph{\underline{B}}_L$. For example, at each iteration, we may update the $i_1^{th}$, $\cdots$,
$i_{n-k+M}^{th}$ LLR's rather than all of them (where $i_1$ through $i_n$ are sorted in ascending reliability).
The number of bits in the reliable part $M$ can be adjusted to control the complexity.

In the computation of the $\tanh$ of each check, we can also make
approximations to reduce the complexity. For instance, min-sum can
be used instead of SPA in Step 3 Algorithm \ref{alg:generic}
\cite{arshad}. Furthermore, since the bit reliabilities are first
ordered, the minimum of the absolute value of the LLR's in the
dense part of the parity check matrix is known. Thus, we can
approximate the $\tanh$ of all the bits in the reliable part using
the $\tanh$ of the minimum value. This modification can
significantly reduce the floating point operation complexity while
retaining most of the performance gain.

More sophisticated updating schemes can also reduce the complexity
of matrix adaptation. El-khamy and McEliece have proposed a scheme
that adapts the parity check matrix from previous ones, which
reduces the overall complexity by $75\%$ (private communication).

\subsection{Symbol-level Adaptation}
\label{subsec:symbol}

Gaussian elimination requires serial update of the rows and is difficult to parallelize. Here we propose an
alternative algorithm that is parallelizable. The idea is to take advantage of the structure of RS codes and
adapt the parity check matrix at the symbol level. Let $S_L = \{i_1,i_2,\ldots,i_{(N-K)} \}$ be a set of $(N-K)$
least reliable symbols (symbol-level reliability can be computed by taking the $\tanh$ product of bit-level
reliabilities or taking the minimum of the bit-level reliabilities). In order to update the parity check matrix
at the symbol level, we need to find a valid parity check matrix for which the submatrix corresponding to the
symbols in $S_L$ is an identity matrix. The detailed procedure is as follows: first, the submatrix corresponding
to the symbols in $S_L$ is filled with an  $(N-K)\times(N-K)$ identity matrix and the rest of the matrix with
unknowns (erasure). The key idea is that computing the unknown symbols in the parity check matrix is equivalent
to finding $(N-K)$ valid codewords of the dual code which will be the rows of the parity check matrix for the
original code. For the $j^{th}$ row, the $i_j^{th}$ entry is 1 and the $i_1^{th}, i_2^{th}, \ldots,
i_{j-1}^{th}, i_{j+1}^{th}, \ldots,i_{N-K}^{th}$ entries are 0s and all other entries are erasures $E$ (i.e.,
all the positions corresponding to the reliable symbols are erased). Since the dual code is an $(N,N-K)$ RS code
with $d_{min}=K+1$ and there are exactly $K$ erasures in each row, Forney's algorithm \cite{wicker_book} can be
used to compute the values in the erased positions. Each decoded codeword corresponds to one row in the original
parity check matrix.  By repeating this procedure for all $(N-K)$ rows, we can get a systematic parity check
matrix over $GF(2^m)$, where the submatrix corresponding to unreliable symbols is the identity matrix. Using the
binary expansion, we can then get the binary parity check matrix and thereafter apply the SPA using it. Unlike
Gaussian elimination, each element in the parity check matrix can be computed independently and, hence, the
whole procedure can be parallelized. This provides a computationally efficient way to obtain a parity check
matrix in the desired form for hardware implementation. Related concepts such as re-encoding have also been used
to reduce the complexity of KV decoding (see \cite{gross_kv}).

\section{Simulation Results}
\label{sec:sim}

In this section, simulation results for iterative decoding of RS codes by adapting the parity check matrix and
its variations over various channel models are presented. The following notations will be used in the legends.
ADP($N_1$,$N_2$) refers to the proposed adaptive decoding scheme. $N_1$ refers to the maximum number of
iterations of iterative decoding. $N_2$ refers to the number of decoding rounds with different groupings of the
unreliable bits (see Section \ref{subsec:grouping}). ADP \& HDD refers to the proposed algorithm with an HDD in
each iteration (see Section \ref{subsec:hdd}). SYM ADP refers to the proposed algorithm with symbol-level
adaptation (see Section \ref{subsec:symbol}). RED(M) ADP refers to the reduced complexity partial updating
schedule with M bits in the reliable part to be updated (see Section \ref{subsec:partial}). MS ADP refers to the
proposed algorithm using ``min-sum'' in updating the LLR's (using min-sum rather than sum-product in Step 3 in
Algorithm \ref{alg:generic} \cite{arshad}). Unless otherwise indicated, all the simulations adopt Deg-2 random
connection (see Section \ref{subsec:deg2}) to improve the asymptotic performance. The damping coefficient
$\alpha$ is also specified on the plots. For comparison, the simulation-based ML lower bounds and analytical ML
upper bounds are also plotted in some figures. The details for obtaining the ML lower bound is described in
\cite{lucas_appdec} and the ML upper bound will be discussed in detail in the following subsection.

To speed up simulation, a genie aided stopping criterion scheme has been used, i.e., the decoding procedure
stops when ADP \& HDD gives the correct codeword. This is mildly optimistic as can be seen from the following
argument. Assume that there is no genie, then the actual decoder will run a fixed number of $N_1$ iterations and
may return a list of codewords (since the HDD may generate different codewords at different iterations). The
actual decoder will pick the most likely codeword from the list. Thus, if the transmitted codeword is the most
likely one, the result of the actual decoder will be the same as that of the genie aided decoder. Only when the
transmitted codeword is not the most likely codeword, i.e., when the ML decoder would have made errors, the
result of the actual decoder may be different from the genie aided decoder and, hence, the genie aided decoder
may be optimistic.


\subsection{Performance of Reed Solomon Codes under Maximum Likelihood Decoding}
\label{subsec:rs_ml}

We first study the performance of RS codes under ML decoding. The
intention is to show that RS codes are themselves good codes (even
for medium rate long codes) and the performance loss is due to the
suboptimal symbol-level bounded distance decoder. The weight
enumerator of an RS code under a specific binary image expansion
is in general unknown. In this paper, we study the performance of
the averaged ensemble of RS codes \cite{retter_gs} under ML
decoding using the Divsalar bound \cite{divsalar}. The averaged
ensemble of the RS code is taken by averaging over all possible
binary expansions. For details, we refer to \cite{retter_gs} and
more recent work \cite{el-khamy_mlbound}.

\begin{figure}
\begin{center}
\includegraphics[width=3.0in]{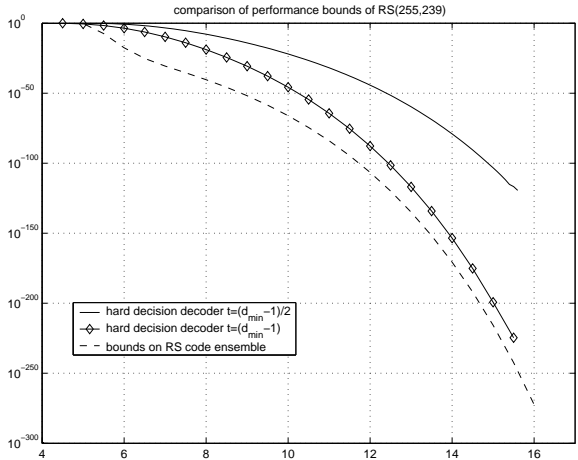}
\caption{Comparison of Performance Bounds of RS(255,239)} \label{fig:rs_bound239}
\end{center}
\end{figure}

We first investigate the performance of a widely used high rate code, i.e., RS(255,239). In Figure
\ref{fig:rs_bound239}, we plot the upper bound on the performance of RS ensemble under ML decoding, HDD with
error correction radius $t = (d_{min}-1)/2$ and a hypothetical decoder which can correct up to $t = (d_{min}-1)$
symbol errors (that is we assume the genie decoder can decode the received vector as long as it is within the
distance of $t = (d_{min}-1)$ at symbol-level from the transmitted codeword). We can see that, the HDD is
asymptotically 3dB worse than the performance under ML decoding (the largest gap is about 4dB, which appears at
around an FER $= 10^{-20}$). The hypothetical decoder is optimal for asymptotically large SNR's. However, this
happens only at very low FER (say, at an FER = $10^{-200}$), which is impractical for most of the applications.
For practical SNR's, there is a loss of approximately 2dB compared to the performance of the ML decoder.

\begin{figure}
\begin{center}
\includegraphics[width=3.0in]{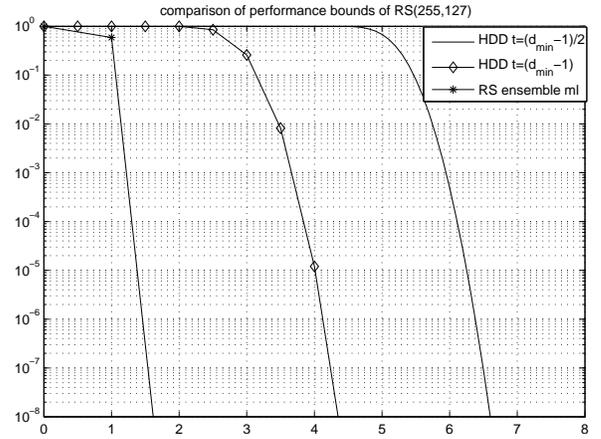}
\caption{Comparison of Performance Bounds of RS(255,127) in Practical SNR Region} \label{fig:rs255127}
\end{center}
\end{figure}

We further investigate a medium rate code RS(255,127) $R = 0.498
\approx 0.5$ in Figure \ref{fig:rs255127}. We can see that the
performance under ML decoding of the RS ensemble reaches an FER =
$10^{-4}$ at an $E_b/N_0 = 1.2$dB and outperforms the hypothetical
decoder and HDD by 2.5dB and 5dB, respectively. The ML performance
of RS ensemble is only 0.6~dB away from the sphere packing bound
\cite{dolinar}, making it comparable to the best known turbo and
LDPC codes. Note that for this code, all known decoders up to now
are still away from the performance under ML decoding, making it
difficult to obtain good simulation based lower bounds to estimate
the ML performance of the RS code.

The above examples show that the symbol-level bounded distance
decoding does not fully exploit the error correction capability of
the code. The hypothetical decoder, which decodes up to $t =
d_{min}-1$, still performs far away from ML decoding, which
suggests that an alternative design principle should be adopted
for RS soft decision decoding. Besides, the analytical performance
bounds of RS codes under ML decoding are of interest as benchmarks
for suboptimal decoders as will be discussed in the following
subsections.

\subsection{AWGN Channels}
\label{subsec:awgn}

\begin{figure}
\begin{center}
\includegraphics[width=3.0in]{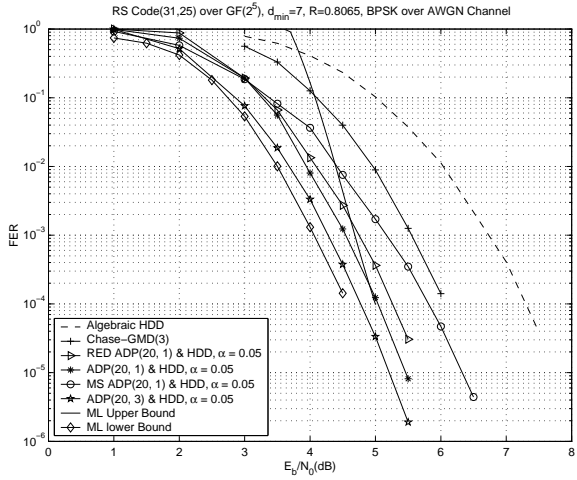}
\caption{RS (31,25) over AWGN channel} \label{fig:rs3125}
\end{center}
\end{figure}

We first present results for the RS (31,25) code over the AWGN channel in Fig. \ref{fig:rs3125}. For this code,
standard belief propagation (BP) decoding (either with or without the damping coefficient, not plotted in the
figure) has little gain (within 0.5~dB from algebraic HDD) due to the large number of short cycles. However, the
proposed ADP(20,1) \& HDD provides a 2.3~dB gain over HDD and an 1.0~dB over Chase-GMD(3) at an FER = $10^{-4}$.
Using the grouping method, the proposed ADP(20,3) \& HDD can approach the ML lower bound within 0.25dB at an FER
= $10^{-4}$. The reduced complexity version RED(20) ADP(20,1) incurs 0.2dB performance loss compared with the
generic ADP and outperforms MS ADP by about 0.5dB at an FER = $3\times10^{-5}$. The ML upper bound over RS
averaged ensemble is also plotted for comparison. It can be seen that the ML upper bound is 0.5dB away from the
ML lower bound at an FER = $10^{-4}$ and these two bounds converge in the high SNR region.

\begin{figure}
\begin{center}
\includegraphics[width=3.0in]{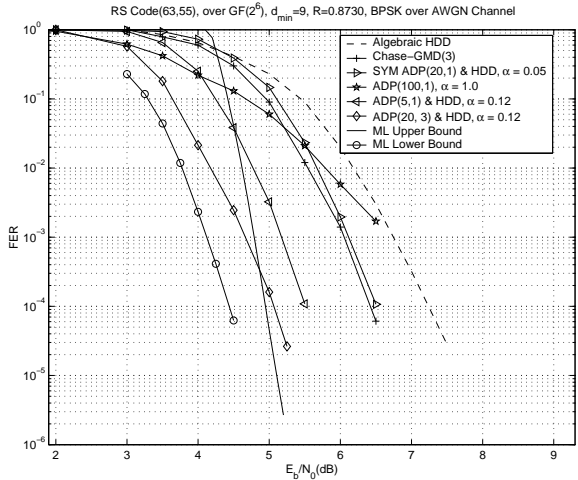}
\caption{RS (63,55) over AWGN channel} \label{fig:rs6355}
\end{center}
\end{figure}

Now we consider the (63,55) RS code. The performance is shown in Fig. \ref{fig:rs6355}. For this code, standard
BP performs even worse than HDD (not plotted in the figure). However, the proposed algorithm ADP(5,1) \& HDD
provides 1.95~dB and 1.05~dB gain over algebraic HDD and Chase-GMD$(3)$ at an FER = $10^{-4}$. ADP(20,3)
performs about 0.7~dB within ML lower bound at an FER = $10^{-4}$. It also provides another 0.3~dB gain over
ADP(5,1). Similar to other gradient descent methods, the damping coefficient of the adaptive algorithm must be
carefully chosen to control the updating step width. The performance curve of ADP(100,1) without damping or
Deg-2 connection has a flat slope and the asymptotic gain diminishes, which is mainly due to the overshooting of
the update scheduling such that the decoder tends to converge to a wrong codeword quickly. SYM ADP(20,1) \& HDD
also provides a non-trivial gain of about 0.7dB over HDD at an FER = $10^{-4}$, which is comparable to
Chase-GMD(3) while the complexity is significantly smaller. The ML upper bound also converges to the ML lower
bound in the high SNR region as in the previous cases.

\begin{figure}
\begin{center}
\includegraphics[width=3.0in]{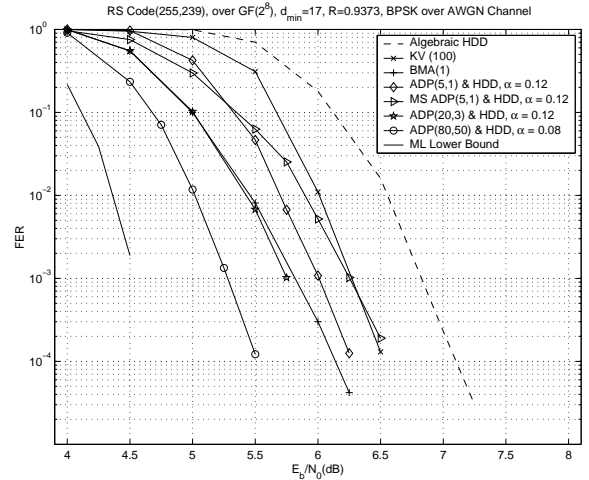}
\caption{RS (255,239) over AWGN channel} \label{fig:rs255239}
\end{center}
\end{figure}

Simulation results for the RS $(255,239)$ code over the AWGN channel are shown in Fig.~\ref{fig:rs255239}. When
large complexity is tolerable, ADP(80, 50) \& HDD outperforms the popular KV method proposed in \cite{gross_kv}
(with maximum multiplicity 100) by 1.0~dB and algebraic HDD by 1.65~dB respectively at an FER = $10^{-4}$. We
also compare this algorithm with BMA order-1 \cite{fossorier_bma}, ADP(80, 50) \& HDD is also about 0.6~dB
better than BMA (1) at an FER = $10^{-4}$. Compared with the ML lower bound obtained by using a near ML decoding
algorithm recently proposed in \cite{fossorier_bias}, the adaptive algorithm is still 0.6dB away from ML lower
bound at an FER = $10^{-3}$. With reasonable complexity, ADP(5,1) \& HDD outperforms the KV(100) at an FER =
$10^{-4}$. Using the ``min sum'' approximation, it will incur about 0.3dB loss at an FER = $10^{-4}$. At the
price of a slight increase in complexity, ADP(20,3) \& HDD can provide comparable performance with BMA(1) at FER
= $10^{-4}$.

\subsection{Rayleigh Fading Channels}
\label{subsec:rayleigh}

\begin{figure}
\begin{center}
\includegraphics[width=3.0in]{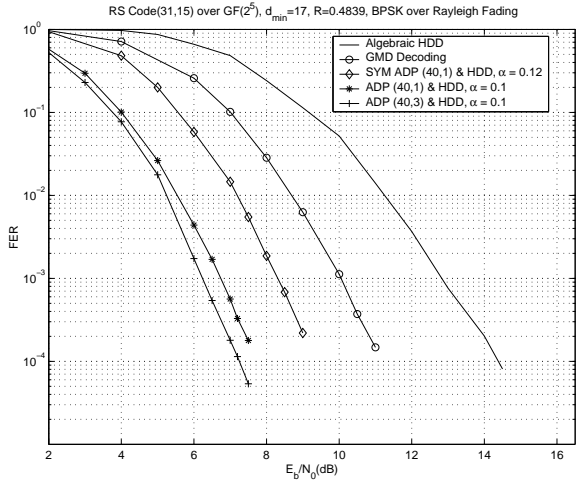}
\caption{RS (31,15) over fully interleaved slow fading channel} \label{fig:rs3115}
\end{center}
\end{figure}

\begin{figure}
\begin{center}
\includegraphics[width=3.0in]{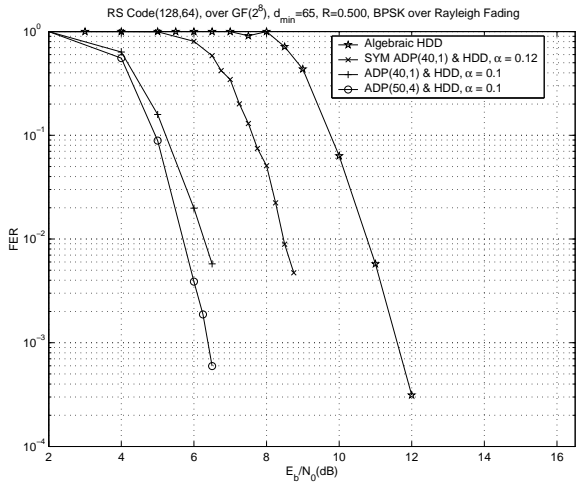}
\caption{RS (128,64) over fully interleaved slow fading channel} \label{fig:rs12864}
\end{center}
\end{figure}

Now we study the performance of the proposed iterative decoding of
RS codes over Rayleigh fading channels. It is assumed that perfect
channel state information is available at the receiver (CSIR). We
first assume BPSK modulation where the coded bits are fully
interleaved at symbol-level, so that fading remains constant over
one symbol but changes from symbol to symbol. The performance of
an RS(31,15) code is shown in Fig. \ref{fig:rs3115}, the proposed
algorithm ADP(40,1) \& HDD outperforms algebraic HDD and GMD
decoding by 6.5~dB and 3.3~dB respectively at an FER = $10^{-4}$.
ADP(40,3) \& HDD can further improve the asymptotic performance.
The performance of  SYM ADP(40,1) \& HDD is also plotted. We see
that it also offers about 5~dB gain over HDD and 1.8~dB gain over
GMD decoding respectively at an FER = $10^{-4}$. Similar results
are observed for long codes with rate $R = 0.5$. The performance
of a shortened RS(128,64) over GF(256) is given in Fig.
\ref{fig:rs12864}. The proposed decoding scheme provides several
dB gain over HDD. This is a nontrivial gain considering the
powerful burst error correction capability of HDD.

\begin{figure}
\begin{center}
\includegraphics[width=3.0in]{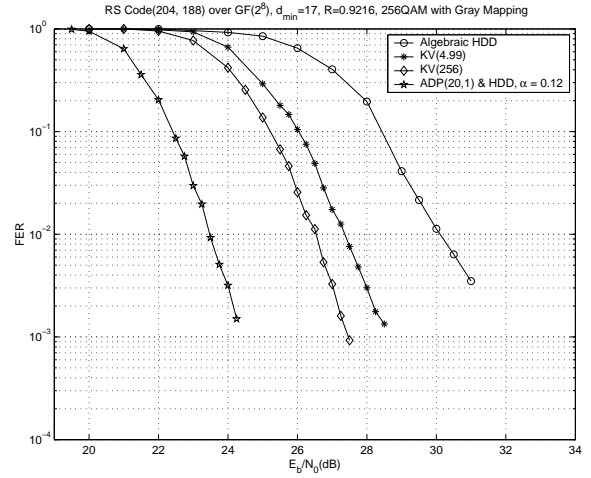}
\caption{RS (204,188) coded modulation over Rayleigh fading channel} \label{fig:rs204188}
\end{center}
\end{figure}

We also study the performance of RS coded modulation system over a symbol-level fully interleaved channel. We
show in Fig. \ref{fig:rs204188} the performance of a shortened RS(204,188) code with 256QAM modulation and gray
mapping, which has similar settings as many existing standards. We can see from the figure that the proposed
algorithm ADP(20,1) \& HDD outperforms algebraic HDD by more than 7dB at an FER = $10^{-3}$. Compared with KV
decoder, there is also a 3 to 4dB gain. Though KV decoder takes the symbol-level soft information directly, its
performance is mainly limited by the algebraic bounded distance decoding kernel.

\section{Discussions and Conclusion}
\label{sec:conclusion}

There are several potential extensions of the adaptive algorithm. Firstly, the gain of the proposed scheme may
diminish at high SNR's for long codes. Further improvement of the generic decoder without significantly
increasing the complexity remains an challenging problem. It is favorable that the structure of the RS codes can
be taken into account in conjunction with the adaptive algorithm. Therefore, Vardy and Be'ery's coset
decomposition \cite{vardy_rs} seems to be a promising way to represent the $\textbf{H}_b$ using a relatively
sparse form. It is also natural to apply some more sophisticated decoding techniques (e.g. constructing some
sub-trellis with reasonable complexity) and adopt the idea of the adaptive algorithm to improve the decoding
performance. Secondly, from our simulation experience, when the channel has memory (say ISI channel or some FSK
signaling), the performance gain of adaptive algorithm (without turbo equalization) diminishes. How to extend
the adaptive scheme to detection and equalization such that they can generate good quality bit-level soft
information is under investigation. Thirdly, asymptotic performance analysis of the adaptive algorithm is also
of interest. Ahmed \emph{et al.} \cite{arshad} showed that using a certain probabilistic model, the performance
of the adaptive algorithm under min-sum approximation can be evaluated using the OSD bound. However, the
performance bounds for the exact scheme is still of interest especially in the high SNR's.

In conclusion, we present a novel iterative SISO decoding algorithm of RS codes by adapting the parity check
matrix. The proposed algorithm can be geometrically interpreted as a two-stage gradient descent algorithm with
an adaptive potential function. Simulation results show that the proposed algorithm compares favorably with
known RS codes soft decoding methods over various channels for a wide range of RS codes of practical interest.
Besides, the proposed algorithm and its variations also provide flexible performance-complexity trade-off for
different applications.

\section*{Acknowledgment}
The authors wish to thank R.~Koetter for pointing out the two-stage optimization essence of the proposed
decoding algorithm. They are grateful to W.~Jin and M.~Fossorier for providing the simulation based ML lower
bound and many other precious suggestions. The authors appreciate the constructive comments from three anonymous
reviewers that have greatly improved the presentation of this paper. J.~Jiang also would like to thank
M.~El-Khamy for many inspiring discussions.


\begin{thebibliography}{10}
\providecommand{\url}[1]{#1} \csname url@rmstyle\endcsname \providecommand{\newblock}{\relax}
\providecommand{\bibinfo}[2]{#2} \providecommand\BIBentrySTDinterwordspacing{\spaceskip=0pt\relax}
\providecommand\BIBentryALTinterwordstretchfactor{4}
\providecommand\BIBentryALTinterwordspacing{\spaceskip=\fontdimen2\font plus
\BIBentryALTinterwordstretchfactor\fontdimen3\font minus
  \fontdimen4\font\relax}
\providecommand\BIBforeignlanguage[2]{{%
\expandafter\ifx\csname l@#1\endcsname\relax
\typeout{** WARNING: IEEEtran.bst: No hyphenation pattern has been}%
\typeout{** loaded for the language `#1'. Using the pattern for}%
\typeout{** the default language instead.}%
\else \language=\csname l@#1\endcsname \fi #2}}

\bibitem{forney_gmd}
G.~D.~F. Jr., ``Generalized minimum distance decoding,'' \emph{IEEE Trans.
  Information Theory}, vol.~12, pp. 125--131, Apr. 1996.

\bibitem{chase_chase}
D.~Chase, ``Class of algorithms for decoding block codes with channel
  measurement information,'' \emph{IEEE Trans. Information Theory}, vol.~18,
  pp. 170--182, Jan. 1972.

\bibitem{tang_cga}
H.~Tang, Y.~Liu, M.~Fossorier, and S.~Lin, ``Combining {Chase}-2 and {GMD}
  decoding algorithms for nonbinary block codes,'' \emph{IEEE Communication
  Letters}, vol.~5, pp. 209--211, May. 2000.

\bibitem{koetter_kv}
R.~Koetter and A.~Vardy, ``Algebraic soft-decision decoding of {Reed}-{Solomon}
  codes,'' \emph{IEEE Trans. Information Theory}, vol.~49, pp. 2809--2825, Nov.
  2003.

\bibitem{el-khamy_kv}
M.~E. Khamy, R.~J. McEliece, and J.~Harel, ``Performance enhancements for
  algebraic soft-decision decoding of {Reed}-{Solomon} codes,'' in \emph{Proc.
  DIMACS}, to be published.

\bibitem{nayak_kv}
N.~Ratnakar and R.~Koetter, ``Exponential error bounds for algebraic
  soft-decision decoding of {Reed}-{Solomon} codes,'' \emph{IEEE Trans.
  Information Theory}, vol.~51, pp. 3899--3917, Nov. 2005.

\bibitem{gross_vlsi}
W.~J. Gross, F.~R. Kschischang, R.~Koetter, and P.~G. Gulak, ``Towards a {VLSI}
  architecture for interpolation-based soft-decision {Reed}-{Solomon}
  decoders,'' \emph{Journal of VLSI Signal Processing}, vol.~39, pp. 93--111,
  Jan.-Feb. 2005.

\bibitem{vardy_rs}
A.~Vardy and Y.~Be'ery, ``Bit-level soft-decision decoding of {Reed}-{Solomon}
  codes,'' \emph{IEEE Trans. Communications}, vol.~39, pp. 440--444, Mar. 1991.

\bibitem{ponn_rs}
V.~Ponnampalam and B.~Vucetic, ``Soft decision decoding of {Reed}-{Solomon}
  codes,'' \emph{IEEE Trans. Communications}, vol.~50, pp. 1758--1768, Nov.
  2002.

\bibitem{fossorier_osd}
M.~P.~C. Fossorier and S.~Lin, ``Soft-decision decoding of linear block codes
  based on ordered statistics,'' \emph{IEEE Trans. Information Theory},
  vol.~41, pp. 1379--1396, Sep. 1995.

\bibitem{yingquan_osd}
Y.~Wu, R.~Koetter, and C.~Hadjicostis, ``Soft-decision decoding of linear block
  codes using preprocessing,'' in \emph{Proc. ISIT}, Chicago, IL, Jul. 2004.

\bibitem{hu_osd}
T.~Hu and S.~Lin, ``An efficient hybrid decoding algorithm for {Reed}-{Solomon}
  codes based on bit reliability,'' \emph{IEEE Trans. Communications}, vol.~51,
  pp. 1073--1081, Jul. 2003.

\bibitem{fossorier_bma}
M.~P.~C. Fossorier and A.~Valembois, ``Reliability-based decoding of
  {Reed}-{Solomon} codes using their binary image,'' \emph{IEEE Communication
  Letters}, vol.~7, pp. 452--454, Jul. 2004.

\bibitem{hagenauer_app}
J.~Hagenauer, E.~Offer, and L.~Papke, ``Iterative decoding of binary block and
  convolutional codes,'' \emph{IEEE Trans. Information Theory}, vol.~42, pp.
  429--445, Mar. 1996.

\bibitem{ungerboeck_rs}
G.~Ungerboeck, ``Iterative soft decoding of {Reed}-{Solomon} codes,'' in
  \emph{Proc. ISTC}, Brest, France, Sep. 2003.

\bibitem{jiang_ssid}
J.~Jiang and K.~R. Narayanan, ``Iterative soft decoding of {Reed} {Solomon}
  codes,'' \emph{IEEE Communication Letters}, vol.~8, pp. 244--246, Apr. 2004.

\bibitem{yedidia_gfft}
J.~S. Yedidia, ``Sparse factor graph representations of {Reed}-{Solomon} and
  related codes,'' in \emph{Proc. ISIT}, Chicago, IL, Jul. 2004.

\bibitem{yedidia_gbp}
J.~S. Yedidia, J.~Chen, and M.~Fossorier, ``Generating code representations
  suitable for belief propagation decoding,'' in \emph{Proc. Allerton},
  Monticello, IL, Oct. 2002.

\bibitem{lin_book}
S.~Lin and D.~J. Costello, \emph{Error Control Coding: Fundamentals and
  Applications}, 1st~ed.\hskip 1em plus 0.5em minus 0.4em\relax Prentice Hall,
  1983.

\bibitem{lucas_appdec}
R.~Lucas, M.~Bossert, and M.~Breitbach, ``On iterative soft-decision decoding
  of linear binary block codes and product codes,'' \emph{IEEE Journal of
  Selected Areas in Communication}, vol.~16, pp. 276--296, Feb. 1998.

\bibitem{farrell_opt}
K.~Farrell, L.~Rudolph, C.~Hartmann, and L.~Nielsen, ``Decoding by local
  optimization (corresp.),'' \emph{IEEE Trans. Information Theory}, vol.~29,
  pp. 740--743, sep. 1983.

\bibitem{gallager63}
R.~Gallager, \emph{Low-Density Parity-Check Codes}.\hskip 1em plus 0.5em minus
  0.4em\relax Cambridge, MA: {MIT} Press, 1963.

\bibitem{pearl98}
R.~J. McEliece, D.~J.~C. MacKay, and J.~F. Cheng., ``Turbo decoding as an
  instance of pearl's ``belief propagation'' algorithm,'' \emph{IEEE Journal on
  Selected Areas in Comm.}, vol.~16, no.~2, pp. 140--152, Feb. 1998.

\bibitem{elias_bec}
P.~Elias, ``Coding for two noisy channels,'' in \emph{In Information Theory,
  Third London Symposium}, 1955.

\bibitem{fossorier_iist}
M.~P.~C. Fossorier, ``Reliability-based soft-decision decoding with iterative
  information set reduction,'' \emph{IEEE Trans. Information Theory}, vol.~48,
  pp. 3101--3106, Dec. 2002.

\bibitem{fossorier_jsac}
------, ``Iterative reliability-based decoding of low-density parity check
  codes,'' \emph{IEEE Journal of Selected Areas in Communication}, vol.~19, pp.
  908--917, May. 2001.

\bibitem{el-khamy_hybrid}
M.~E. Khamy and R.~J. McEliece, ``Iterative algebraic soft decision decoding of
  {Reed}-{Solomon} codes,'' in \emph{Proc. ISITA}, Parma, Italy, Mar. 2004, pp.
  1456 -- 1461.

\bibitem{arshad}
A.~Ahmed, R.~Koetter, and N.~R. Shanbhag, ``Performance analysis of the
  adaptive parity check matrix based soft-decision decoding algorithm,'' in
  \emph{Proc. Asilomar Conf. Signals, Systems, and Computers}, Pacifi c Grove,
  CA, Nov. 2004.

\bibitem{wicker_book}
S.~B. Wicker, \emph{Error Control Systems for Digital Communication and
  Storage}, 1st~ed.\hskip 1em plus 0.5em minus 0.4em\relax Prentice Hall, Jan
  1995.

\bibitem{gross_kv}
W.~J. Gross, F.~R. Kschischang, R.~Koetter, and P.~G. Gulak, ``Applications of
  algebraic soft-decision decoding of {Reed}-{Solomon} codes,'' \emph{IEEE
  Trans. Communication}, submitted 2003.

\bibitem{retter_gs}
C.~Retter, ``The average weight-distance enumerator for binary expansions of
  {Reed}-{Solomon} codes,'' \emph{IEEE Trans. Information Theory}, vol.~48, pp.
  1195--1200, Mar. 2002.

\bibitem{divsalar}
D.~Divsalar, ``A simple tight bound on error probability of block codes with
  application to turbo codes,'' TMO Progress Report, TR 42-139, Nov. 1999.

\bibitem{el-khamy_mlbound}
M.~E. Khamy and R.~J. McEliece, ``Bounds on the average binary minimum distance
  and the maximum likelihood performance of {Reed}-{Solomon} codes,'' in
  \emph{Proc. Allerton}, Monticello, IL, Oct. 2004.

\bibitem{dolinar}
S.~Dolinar, D.~Divsalar, and F.~Pollara, ``Code performance as a function of
  block size,'' TMO Progress Report, TR 42-133, May. 1998.

\bibitem{fossorier_bias}
W.~Jin and M.~P.~C. Fossorier, ``Iterative biased reliability-based decoding of
  binary linear codes,'' in \emph{Proc. IEICE and SITA Joint Conference on
  Info. Theory}, Honolulu, HW, May. 2005.

\end{thebibliography}

\begin{biography}
{Jing Jiang} (S'02) received the B.S. degree from Shanghai Jiao Tong University in 2002 and is currently working
toward the M.S./Ph.D. degree at Texas A \& M University in the department of electrical and computer
engineering. His general research interests lie in the areas of channel coding, signal processing and
information theory for wireless communication and digital storage channels.
\end{biography}
\begin{biography}
{Krishna R. Narayanan} (S'92-M'98) received the Ph.D. degree in electrical engineering from Georgia Institute of
Technology in 1998 and is currently an associate professor in the department of electrical and computer
engineering at Texas A~\&~M University. His research interests are in coding theory with applications to
wireless communications and digital magnetic recording and joint source-channel coding. He currently serves as
an editor for the IEEE Transactions on Communications.
\end{biography}

\end{document}